\documentclass{comjnl}

\usepackage{graphicx}
\usepackage{color}
\usepackage[usenames,dvipsnames,svgnames,table]{xcolor}
\usepackage{amsfonts}
\usepackage{booktabs}

\newcommand{\tY}{\checkmark}
\newcommand{\tN}{$\times$}

\graphicspath{{./images/}}

\title[Identifying Compiler Options to Minimise Energy Consumption for Embedded Platforms]{Identifying Compiler Options to Minimise Energy Consumption for Embedded Platforms}

\author{James Pallister}
\affiliation{Department of Computer Science, University of Bristol, Merchant Venturers Building, Woodland Road, Bristol, BS8 1UB,
United Kingdom.}
\email{james.pallister@bristol.ac.uk}

\author{Simon Hollis}
\affiliation{Department of Computer Science, University of Bristol, Merchant Venturers Building, Woodland Road, Bristol, BS8 1UB,
United Kingdom.}

\author{Jeremy Bennett}
\affiliation{Embecosm, Palamos House \#104, 66/67 High Street, Lymington, SO41 9AL, United Kingdom.}

\shortauthors{J. Pallister and S. Hollis and J. Bennett}

\begin{document}

\raggedbottom

\begin{abstract}

This paper presents an analysis of the energy consumption of an extensive number of the optimisations a modern compiler can perform. Using GCC as a test case, we evaluate a set of ten carefully selected benchmarks for five different embedded platforms.

A fractional factorial design is used to systematically explore the large optimisation space ($2^{82}$ possible combinations), whilst still accurately determining the effects of optimisations and optimisation combinations. Hardware power measurements on each platform are taken to ensure all architectural effects on the energy consumption are captured.

We show that fractional factorial design can find more optimal combinations than relying on built in compiler settings. We explore the relationship between run-time and energy consumption, and identify scenarios where they are and are not correlated.

A further conclusion of this study is the structure of the benchmark has a larger effect than the hardware architecture on whether the optimisation will be effective, and that no single optimisation is universally beneficial for execution time or energy consumption.

\end{abstract}

\maketitle

\section{Introduction}

\begin{table*}
	\centering
	\begin{tabular}{l l l l l}
		\textbf{Processor} & \textbf{Board name}  & \textbf{RAM} & \textbf{$\!\!\!\!\!\!\!\!\!$Core clock} & \textbf{Other} \\
		\midrule
		ARM Cortex-M0 	 & STM32F0DISCOVERY		& 8KB		& 48 MHz		  & 64KB Flash\\
		ARM Cortex-M3 	 & STM32VLDISCOVERY		& 8KB		& 24 MHz		  & 128KB Flash\\
		ARM Cortex-A8 	 & BeagleBone			& 256MB		& 500 MHz		  & VFP/NEON, superscalar\\
		Adapteva Epiphany 	 & EMEK3				& 32KB/core & 400 MHz		  & FPU,superscalar,16 core NoC\\
		XMOS L1 		 & XK1					& 64KB		& 100 MHz 		& 4$\times$100MHz hardware threads \\
	\end{tabular}
	\caption{The platforms explored in this paper along with some relevant details.}
	\label{Table:Platforms}
\end{table*}

Energy consumption is rapidly becoming one of the most important design constraints when writing software for embedded platforms. In the hardware space there are many features, such as clock gating and dynamic frequency and voltage scaling, to reduce the power consumption of electronic devices. However, inefficient software can negate any gains from the hardware, so the combination of software and hardware must be considered together when exploring energy usage. This study focuses on processors for embedded platforms, because energy efficiency is particularly important for many of their target applications.

Optimising the software for low energy consumption is particularly important when adhering to a strict power budget. This is the case in many deeply embedded systems. In these devices the processor is a significant consumer of energy --- a previous study characterised the CPU power usage of a handheld device to be between 20 and 40\% of the total system power~\cite{SmartPhonePower}. A further study, based on 45~nm technology data from~\cite{Lofikamran2012}, calculated the power dissipation of the processors in a 64-core network on chip to be 40\% of the entire system~\cite{Hollis2012}. This category was the largest, ahead of memory, network, and I/O.

Compiler optimisations have the potential for energy savings with no changes to existing hardware or software --- just tweaking the compiler's parameters can have a large effect on the energy consumption~\cite{Pan2006a}. Sometimes this manifests in spikes of higher power followed by longer periods of lower power; at other times the power is maintained at a lower level. Both can reduce total energy. This relationship is complex, with the program, processor architecture and specific compiler options interacting. Furthermore, different optimisation passes interact with each other, so an option's efficacy cannot be tested in isolation. For example, inlining a function may mean that more effective common subexpression elimination can be performed, increasing the performance more than either option individually. Many approaches have attempted to solve this optimisation selection problem, using techniques such as statistical methods~\cite{Haneda2005}, genetic algorithms~\cite{Lin2008} and iterative compilation~\cite{Kisuki1999}. All of these studies conclude that performance can be increased by choosing the correct set of optimisations, but exploring the space to find this set is challenging.

The energy an application takes can be measured using a set-up as shown in Fig.~\ref{Fig:setup}. A shunt resistor is inserted between the power supply and processor, allowing the voltage drop across it to be measured and amplified. This can be converted into an instantaneous power reading by considering the resistance of the shunt. The power logger assigns a timestamp to each power sample, allowing them to be integrated into a total amount of energy consumed during the execution of the application.

The following section covers the overall aims and hypotheses we wish to address in this paper. Then, the related work is discussed. Following this section, our approach to the problem of benchmark selection and compiler flags is given. The initial high-level results are presented with discussion of the first two hypotheses in Sect.~\ref{sect:time_and_energy}. In Sect.~\ref{sec:ffd}, there is a short introduction to fractional factorial design, followed by the results obtained using this technique. The case studies of the most effective optimisations, and the interactions between optimisations is given in Sect.~\ref{sect:universality_flags}. Finally, concluding remarks to the application developer and the compiler writer are made.

\section{Overview of this Work}

\begin{figure}
	\includegraphics[width=\linewidth]{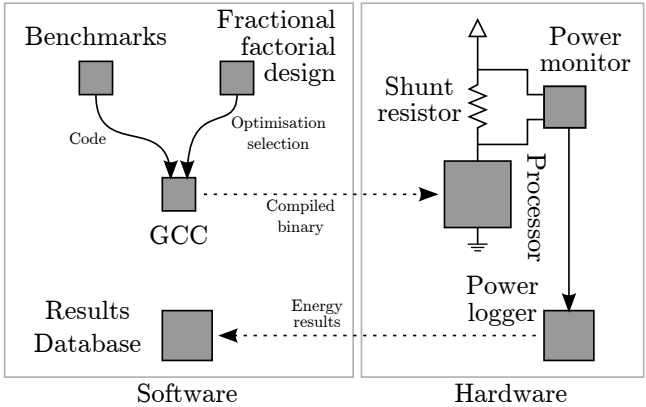}
	\caption{The hardware and software setup used to take the measurements.}
	\label{Fig:setup}
\end{figure}

The overall aim of this work is to identify compiler optimisations that are effective at reducing a benchmark's energy consumption. This is accomplished by using fractional factorial design to account for interactions between the optimisations, without having to enumerate all combinations of optimisations. This analysis is performed for multiple benchmarks and platforms, allowing general conclusions to be drawn about how the optimisations affect energy consumption.

We investigate the following hypotheses:
\begin{enumerate}
	\item The time and energy required for a computation are always proportional to one another. We find investigate and present examples where energy and time are not correlated and explain why (Sect.~\ref{sect:time_and_energy}).
	\item There exists a set of compiler optimisations that gives a lower energy consumption than the predefined optimisation levels (Sect.~\ref{sec:optimisation_potential}).
	\item It is possible to search the compiler optimisation space in an efficient and systematic manner, to assign each optimisation an overall effectiveness (Sect.~\ref{sec:ffd}).
	\item There is no universally good optimisation across multiple benchmarks and platforms (Sect.~\ref{sect:universality_flags}).
\end{enumerate}

We will evaluate the validity of these hypotheses by performing a series of practical experiments which target the set of optimisations enabled at various optimisation levels of a real compiler. This allows the optimisations to be measured for their effect on energy. Three types of experiment are performed in this study:

\begin{description}
	\item[High-level.] Each optimisation level (predefined set of optimisations) is tested for each benchmark and platform. This is a small number of tests, exploring the overall effect of the optimisation levels.
	\item[Fractional factorial design.] A fractional factorial design is used to find the effectiveness of each optimisation flag defined at the optimisation level. This experiment is repeated for each optimisation level, and for each benchmark and platform combination. A large number of tests are performed for each combination of benchmark, platform and optimisation level. This is the first time this technique has been applied across multiple platforms for the purpose of analysing compiler optimisations.
	\item[Case study.] Two case studies are performed. The first looks at the most effective optimisation flags across benchmarks and platforms, extracted from the fractional factorial design. The second explores the interactions between optimisations by exhaustively enumerating every combination of a small set of optimisations.
\end{description}

All the energy measurements in this paper are taken using physical measurement circuitry attached to the processors. This avoids the use of models which could be inaccurate, or modelling synthetic processors with no real world counterpart. A diagram of the software and hardware setup is shown in Fig.~\ref{Fig:setup}. By using commonly available platforms and processors, along with some more novel architectures, the results are applicable in general while still providing insight into how different types of architectures perform. It is important to consider a wide range of platforms when analysing energy consumption, since the structure of the processor's pipeline has a large effect on both the conditions in which energy consumption is high, and the types of optimisations which are effective. The platforms examined, shown in Tab.~\ref{Table:Platforms}, have a wide range of pipeline depths, memory bandwidth capabilities and numbers of registers, allowing analysis of how an optimisation's effect on energy changes under these circumstances (an analysis of how the processor's features affect the energy consumption is given in Sect.~\ref{sect:time_and_energy}).

This work makes the following contributions:

\begin{itemize}
	\item The use of fractional factorial design to analyse a previously intractable optimisation space of GCC 4.7's optimisation options.
	\item Analysis of relative importance of each optimisation across multiple benchmarks and platforms.
	\item The answers to the previously given hypotheses.
	\item Commentary on how these techniques and results can be used by application developers and compiler writers.
\end{itemize}

\section{Related Work}

\subsection{Compilers \& Energy}

To date there has been very little work that extensively explores the effect that different compiler optimisations have on energy consumption. However, there have been many studies that look at the effect of optimisations on execution time~\cite{Haneda2005,Purini2013}, and several studies suggesting that execution time can be used as a proxy for energy usage~\cite{EffectOfCompilerOptimisationsOnPentium4, Ibrahim2009b}.

The topic of performance and energy being highly correlated is addressed in~\cite{CompilingForPerformancePower}. This work explored several different overall optimisation levels, as well as four specific optimisations, using the \textit{Wattch simulator}~\cite{Wattch} to estimate energy results. However, the specific optimisations were all applied individually on top of the first optimisation level, without exploring any possible interactions between the optimisations. The main conclusion drawn from this study was that most optimisations reduce the number of instructions executed, hence reducing energy consumption and execution time simultaneously.

Of the studies that look at individual optimisations and their effects on energy or power, most focus on only a few optimisations in isolation and few consider multiple platforms with different architectural features. Commonly explored optimisations, such as loop unrolling~\cite{Ayala2004}, loop fusion~\cite{Zhu2004}, function inlining~\cite{Kim2012} and instruction scheduling~\cite{Toburen1998}, have been examined extensively for different platforms using both simulators and hardware measurements.

A drawback of those studies that explore energy consumption is that many of them choose to use simulators as opposed to taking hardware measurements. The Wattch simulator is designed to allow easy energy measurements while exploring architectural configurations and is established at being within 10\% of an industry layout-level power tool. However, Wattch does not model every hardware component in the processor, which makes it difficult to be certain about the total energy consumption of the processor.

SimplePower~\cite{SimplePower} is another simulator that has been used to explore the energy consumption of the software running on a processor. This simulator targets a five stage RISC pipeline, with energy consumption estimates based on the number of transitions on bus signal lines as well as various other components.

Various other models have been created to simulate power consumption of the processor, including complex instruction level models~\cite{Steinke2001}, function-level models~\cite{Qu2000} and hybrids of these~\cite{Blume2007}. However, these all suffer the drawback that some energy consumption effects may not be modelled, potentially skewing the results.

\subsection{Optimisations Targeting Energy}
\label{subsect:energyoptimisations}

Many previous studies look at how to utilise \textit{existing} optimisations to target energy consumption. However, all of these optimisations were written with the aim of reducing execution time, not energy consumption. Several other techniques have been proposed to develop optimisations that specifically target energy consumption.

An analysis of the techniques the compiler can perform to optimise for energy was carried out by Tiwari, Malik and Wolfe~\cite{CompilationTechniquesForLowEnergy}. They identified several possible techniques that compilers could use to reduce the energy consumption of programs. They were:
\begin{itemize}
	\item Reorder instructions to reduce switching.
	\item Reduce switching on address lines.
	\item Reduce memory accesses.
	\item Improve cache hits.
	\item Improve page hits.
\end{itemize}

The last three will also normally increase performance as well as reduce energy.

Several novel types of compiler optimisations have been proposed. Seth et al~\cite{Seth2001} explored the possibility of using the compiler to insert \texttt{idle} instructions automatically, increasing the execution time up to a set limit. Using the SIMD pipeline has been shown to decrease energy consumption~\cite{Ibrahim2009} by roughly 25\%. Scheduling instructions to minimise the inter-instruction energy cost was evaluated to be another effective method to reduce a program's energy consumption~\cite{Parikh}. Exploiting differences in energy consumption between other function units has been suggested in~\cite{WhatCanAPoorCompilerDo}, where it is noted that strength reduction may use a more efficient shifter rather than a power hungry multiplier.

The use of scratchpad memory has been used to increase the energy efficiency of processors with these features in~\cite{Steinke}. Other techniques have also been employed to reduce the energy cost of going to memory by accounting for the bit-width required by the variable being accessed~\cite{Cao2001}.

\subsection{Optimisation Choice}

The challenge of choosing the optimisations and their order has been explored and many methodologies proposed for choosing an optimal set of optimisations.

Chakrapani et al. attempt to classify optimisations by the effect they have on performance and energy~\cite{WhatCanAPoorCompilerDo}. This work used both hardware measurements and a gate-level simulation to derive the results, separating the optimisations into the following three classes:
\begin{itemize}
	\item \textbf{Reduction in energy consumption due to increase in performance.} Optimisations in this class reduce the number of cycles or instructions needed to complete the application and thus less overall work is done.
	\item \textbf{Optimisations that reduce energy without improving the performance.} These optimisations reduce the instantaneous power in portions of the program without increasing the number of cycles to compute the result. Scheduling instructions to reduce switching and making use of power efficient functional units often fall into this category.
	\item \textbf{Optimisations that increase energy consumption or decrease performance.} These include optimisations which can sometimes have unexpected performance hits, such as loop unrolling and function inlining. The increase in energy consumption can come from either a longer running time at the same average power, or a higher average power.
\end{itemize}

Iterative compilation has been examined as a possibility for choosing optimisations that reduce power by Gheorghita et al~\cite{IterativeCompilationForEnergy}. In this paper, the effect of different loop unrolling and loop tiling parameters on energy consumption is examined for three linear-algebra-based benchmarks using a simulator. The paper concluded that iterative compilation was an effective method of decreasing energy consumption as well as improving performance.

Other approaches have looked at genetic algorithms for optimisation selection~\cite{Lin2008} and optimisation phase ordering~\cite{Almagor2004}. While these techniques are shown to be effective, they have the drawback that the reasons behind an optimisation's selection is not obvious. Another study~\cite{Lokuciejewski2011} has explored genetic algorithms in the context of optimising multiple objectives. This study used the technique to balance the trade-off between code size, performance and worst-case performance, and could also optimise for any single objective at the cost of the others.

These techniques do not expose the relationships between optimisations, instead opting to search though the optimisation space and making a best guess about where to look next. In this paper, we improve these shortcomings by using fractional factorial design~\cite{BoxHunter} to explore the most effective optimisations and the interactions between them.

Fractional factorial design, as a method for exploring the interactions of compiler optimisations was discussed in~\cite{IntelPaper}. Nine optimisations were examined, using a fractional factorial technique to isolate the interactions and choose a set of optimisations that gave better performance than just enabling all the optimisations. This concept was extended by Haneda et al.~\cite{Haneda2005b} to combine the results of running fractional factorial design on several benchmarks into one set of flags.

A similar study conducted by Patyk et al~\cite{EnergyReductionCompilerOptions}, extended this work to energy efficiency. The study explored a range of GCC's options, with an aim to reduce energy consumption by identifying significant optimisations, then excluding them from further exploration using fractional factorial design. We use this technique to analyse the optimisations rather than optimise for the energy consumption.

These methods all require testing over many different compilations, which is a significant overhead when finding an optimal set. The MILEPOST GCC~\cite{Fursin2011} study implemented an alternative to this, using machine learning to guess which optimisation flags would best apply to a given program. Features are extracted from the program, which are then matched against previous known results from previous compilations. This allows a set of optimisations to be predicted from just the source code. The drawback of this approach is that a large number of programs and optimisation combinations must be used to train the database, and there is not yet knowledge about the appropriate program features for predicting energy consumption or if they exist.

The majority of the studies listed in this section only examine one platform, and it is currently unknown whether their results would apply across several different platforms. Furthermore, iterative compilation~\cite{Kisuki1999} and other adaptive techniques used can leave holes of potential combinations of optimisations unexplored (due to the huge numbers of combinations possible). This can lead to the most optimal configurations not being found.

\section{Approach}

\begin{table}
	\centering
	\begin{tabular}{l l l}
	\textbf{Name}			& \textbf{Source} 	& \textbf{Category} \\
	\midrule
	2D FIR					& WCET 		& Branching, FP	\\
	Blowfish				& MiBench 	& Integer	\\
	CRC32					& MiBench 	& Branching, integer	\\
	Cubic solver			& MiBench 	& Memory, integer	\\
	Dijkstra				& MiBench 	& Branching, integer	\\
	FDCT					& WCET 		& Branching, memory	\\
	Float matmult			& WCET 		& Memory, FP	\\
	Int matmult				& WCET	 	& Memory, integer	\\
	Rjindael				& MiBench 	& Branching, integer	\\
	SHA						& MiBench 	& Memory, integer	\\
	\end{tabular}
	\caption{Benchmarks selected, and the types of instructions they execute intensively (FP short for Floating Point). }
	\label{Table:Benchmarks}
\end{table}

In this paper, we present an improved technique for testing the effectiveness of large numbers of compiler optimisations and their impact on energy consumption and run-times. The technique is based on the concept of fractional factorial design (see Sect.~\ref{sec:ffd}).

\begin{figure*}[tb!]
	\centering
	\includegraphics[width=\textwidth,clip, trim=0cm 0cm 0 2.1cm]{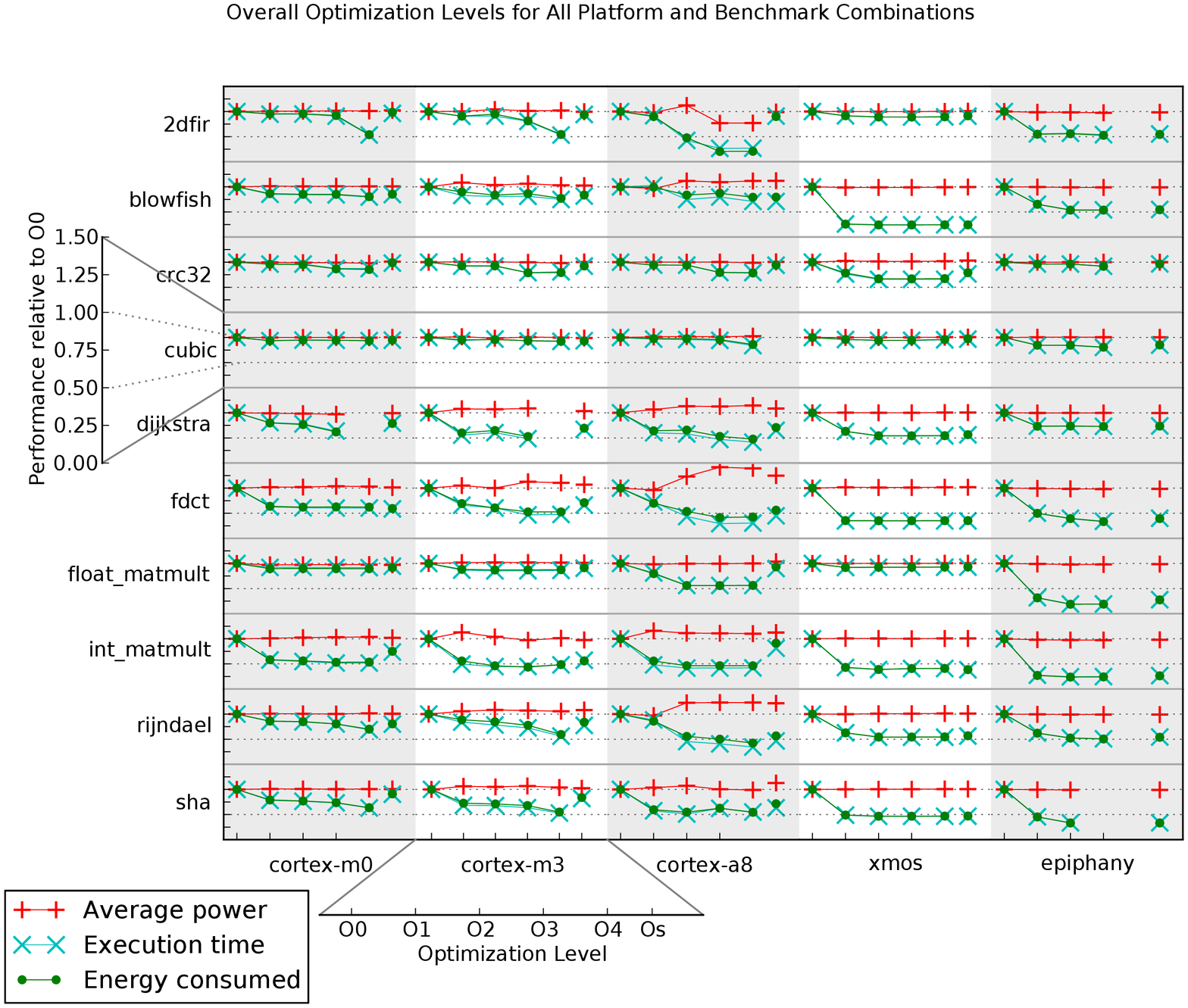}
	\caption{Energy, time and power results for benchmark-platform combinations. Optimisation levels \texttt{O0} to \texttt{O4}. \texttt{O4} is \texttt{O3} with link-time optimisation. The last point is \texttt{Os} --- optimise for space. Some results are unavailable for when the compiler crashed while producing the output binary.}
	\label{Fig:OverallView}
\end{figure*}

\subsection{A New Benchmark Set}

To explore the impact of the optimisations, a realistic set of test input programs is required. There have been many attempts to find representative programs, but none apply to a wide range of embedded processing systems. Frequently a benchmark will require host operating system support, requiring features which are rarely available on deeply embedded platforms. Also the size of the benchmark and the dataset it uses becomes of critical importance when running on platforms with such limited memory. Because of the lack of suitable suites, we evaluated each benchmark from a large number of contemporary suites. Individual benchmarks were considered for inclusion based on their distribution of instruction types. Further details on this are given below, and described in~\cite{BenchmarkPaper}.

This set of 10 benchmarks, shown in Tab.~\ref{Table:Benchmarks}, covers real world and synthetic applications across different aspects of the target platform. These are selected from MiBench~\cite{MiBench} and the Worst Case Execution Time (WCET)~\cite{WCET} suites. Previous work on modelling the energy consumption of processors has shown that the pipelines and functional units enabled have a significant impact on the energy consumption. To cover these points, the benchmarks were characterised according to the following coarse criteria:
\begin{itemize}
	\item \textbf{Integer pipeline intensity.} The frequency at which integer arithmetic instructions occur.
	\item \textbf{Floating point intensity.} The frequency of floating point operations.
	\item \textbf{Memory intensity.} Whether the program requires a large amount of memory bandwidth or not.
	\item \textbf{Branch frequency.} How often the code branches.
\end{itemize}


Similar categories of instruction types have been used previously to give a high level overview of the type of computation an application is performing~\cite{Hennessy2012}. Our categories group similar instruction, such as the loads and stores in MiBench, since energy consumption is predominately related to the target functional unit, rather than the specific operation.

This set of benchmarks is chosen because they do not require a host operating system. This prevents the benchmark from being pre-empted by another process and reducing the accuracy of the results. It also makes the execution of the benchmarks deterministic. For the same reasons, the benchmarks do not perform any I/O.

The benchmarks are also chosen carefully with regards to memory requirements.
The benchmarks are designed to fit into the memory footprint of a wide range of embedded systems, with or without a memory hierarchy. In the cache-based systems we explore, this often has the effect that the benchmark fits entirely into a single level of cache, reducing complexity but potentially also accuracy. This is a trade-off that we have found necessary when generating benchmarks that cover a wide range of hardware platforms, and has the benefit that the benchmarks exhibit predictable memory accesses on most platforms.

\subsection{Compiler Flags}

We explore the impact of compiler optimisations using the GCC toolchain on the architectures shown in Tab.~\ref{Table:Platforms}. GCC exposes its various optimisations via a number of flags that can be passed to the compiler~\cite{GCC}. We explore which flags have a significant impact on energy consumption and execution time.

The experiments are performed with different benchmarks, so a complete picture of architecture, optimisation and application can be seen. Using this combination, the following points of interest can be explored:
\begin{itemize}
 	\item The relationship between time and energy for our benchmarks;
 	\item Architectural effects on energy consumption; and
 	\item Application effects on energy consumption.
 \end{itemize}

By using the techniques we have just outlined, we can rigorously evaluate our hypotheses, answering questions about the relationship between time and energy, and optimisation choice.

\section{Time and Energy}
\label{sect:time_and_energy}

\begin{figure*}
	\includegraphics[width=\linewidth,clip,trim=0.5cm 0.1cm 0cm 1.8cm]{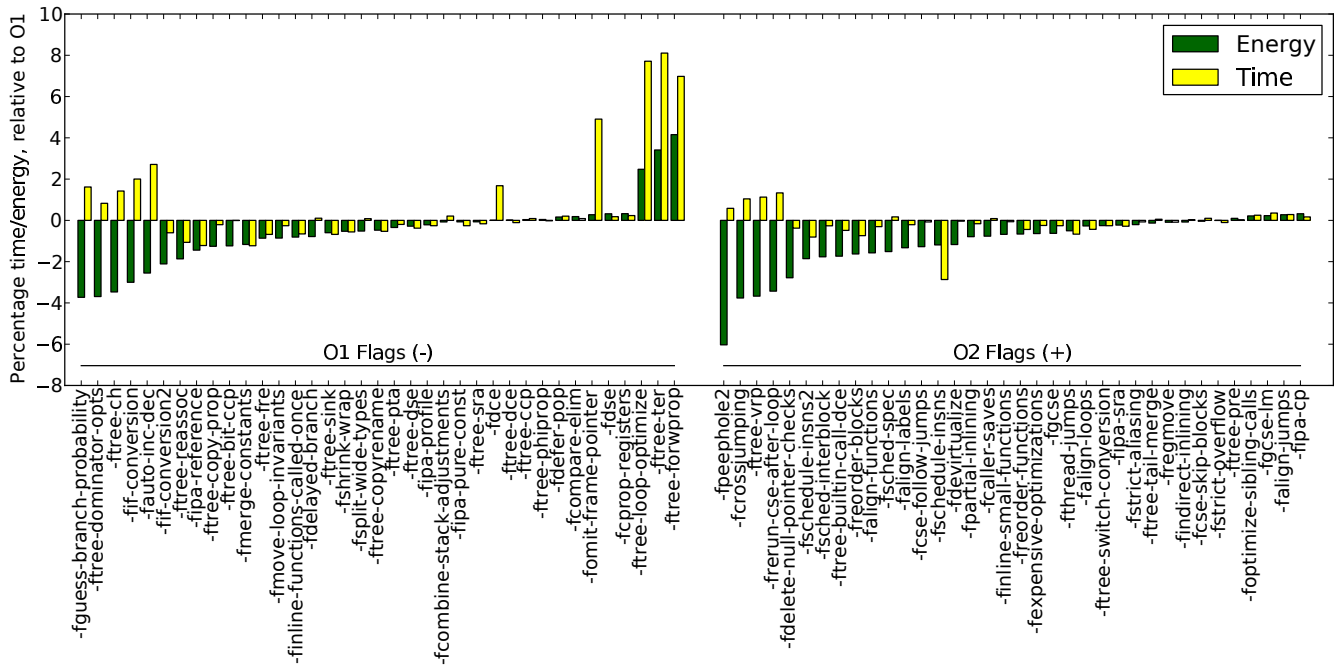}
	\caption{Blowfish benchmark on the Cortex-M3 platform. Individual options are enabled or disabled on top of the \texttt{O1} optimisation level.}
	\label{Fig:AddsubO1Blowfish}
\end{figure*}

The following section addresses the first hypothesis, and show that energy consumption and execution time are proportional to each other across all the benchmarks and platforms. A high level overview of each platform and benchmark for the different optimisation levels is given in Fig.~\ref{Fig:OverallView}. This figure shows a line graph for each combination, displaying the effect of the broad optimisation levels \texttt{O1}, \texttt{O2}, \texttt{O3}, \texttt{O4} (defined as \texttt{O3} with link time optimisation) and \texttt{Os} (optimise for space) on time, energy and average power when compared to the same program with all optimisations disabled.

For the Cortex-M0, very little difference between energy and time is seen due to it being the simplest processor tested, it has a three stage pipeline without forwarding logic. The pipeline behaviour is simple, only stalling if it encounters a load or a branch, thus it is not sensitive to specific code sequences. The Cortex-M3 exhibits very similar behaviour, with some very slight differences between energy and time. The micro-architecture in this processor is more complex, featuring branch speculation and a larger instruction set~\cite{Yiu2010}.

The XMOS processor has a four stage pipeline, similar to the Cortex-M3 in complexity and performance. It should also be noted that the compiler for the XMOS processor uses an LLVM backend~\cite{LLVM} for code generation, featuring different optimisations. Due to this the result set for this processor is not as extensive as the other four, but is still broadly comparable.

The Epiphany processor also sees a large correlation between the energy consumption and execution time. There is some divergence when the superscalar core in the processor is able to dispatch multiple instructions simultaneously. This gives the compiler more potential for creating advantageous code sequences.

The greatest difference between energy and time was discovered while using the Cortex-A8. For the majority of the benchmarks the execution time reduces more than the energy. This is due to multiple instructions being executed simultaneously by the superscalar core, reducing the amount of time taken but not the energy consumption, as the same total work is still being done. We infer from this that the amount of pipeline activity has a significant measurable effect on the energy consumption. The gap is also seen to widen at the \texttt{O2} level, due to instruction scheduling being enabled there.

These results support our first hypothesis that time and energy are broadly correlated. The strongest correlation occurs in the qualitatively `simplest' pipelines. Increasing pipeline complexity means there are more opportunities for architectural energy saving measures (clock gating, etc.) making the complex processor's energy profile more variable and improving the potential for compiler optimisation impact.

\section{Optimisation Potential}
\label{sec:optimisation_potential}
The second hypothesis to explore is that it was possible to find a set of optimisations that perform better than the standard optimisation levels. Fig.~\ref{Fig:AddsubO1Blowfish} shows each option in \texttt{O1} and \texttt{O2} optimisation levels enabled on top of the flags in \texttt{O1}. By examining the left of the graph, it can be seen that by disabling \texttt{-fguess-branch-probability} (in this specific run) the energy decreases by 4\% at the expense of some additional run-time. This shows that a set of optimisations that performs better than the predefined \texttt{O1} optimisation level.

This conclusion is in line with much of the related work, that has focused on choosing a set of optimisations which is more optimal than the standard optimisation levels for a given benchmark.

\section{Fractional Factorial Design}
\label{sec:ffd}

\begin{figure}[bt]
	\centering
	\includegraphics[width=0.9\linewidth]{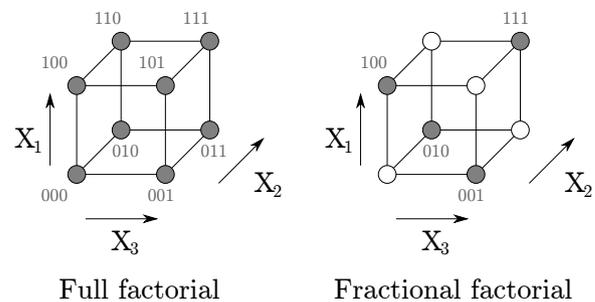}
	\caption{Reducing a 3-factor full factorial design to a `half fraction' design.}
	\label{Fig:FFDTut}
\end{figure}

This section explores the third hypothesis --- a method to systematically explore the optimisation space.

GCC has over 150 different options that can be enabled to control optimisations. The majority of these options are binary --- the optimisation pass is either enabled or disabled. To further complicate matters, an optimisation path may be affected by other passes happening before it. It is not feasible to test all possible combinations of options, therefore a trade-off has to be made. One of our main contributions is to deploy fractional factorial design~\cite{BoxHunter} (FFD) to massively reduce the number of tests to explore the space, whilst still identifying the options that contribute to run-time and energy. This approach has been explored on a small scale in ~\cite{IntelPaper}, where nine optimisations were explored in just 35 tests as opposed to the 512 required for a full factorial design. It has also been explored by Haneda et al. in~\cite{Haneda2005b}, where a fractional factorial design is used to inform the choice of optimizations. We apply this technique to allow us to analyse and draw conclusions about these large number of optimizations.

An example \textit{full} factorial design is shown on the left of Fig.~\ref{Fig:FFDTut}. This example shows three factors with every possible combination enumerated. A fractional factorial design with the number of tests halved is shown on the right, yet still allows the difference between any two factors to be estimated.

The drawback to this approach is that the high-order interactions between options (effects due to multiple options being enabled) will not be discernible. Fortunately, this is not usually a problem as these types of interactions are statistically rare. The degree to which this happens is specified by the FFD's resolution. A {\em resolution 5} design ensures that the main effects are not aliased with anything lower than 4th order interactions.

Using the Yates algorithm~\cite{BoxHunter}, the effect for any single or combination of factors can be found from the data. This gives an estimate for how much this factor or interaction affects the result of the experiment. The Mann-Whitney statistical test is used to determine whether the factor represents a significant change in performance as detailed in~\cite{EnergyReductionCompilerOptions} and~\cite{Haneda2005}.

All FFDs used were generated by the statistical program, R~\cite{R} (a statistical programming language), using the FrF2 library~\cite{FrF2}.

\subsection{FFD Results}
\begin{figure}[t!]
	\includegraphics[width=\linewidth,clip,trim=0.5cm 0 2cm 1.8cm]{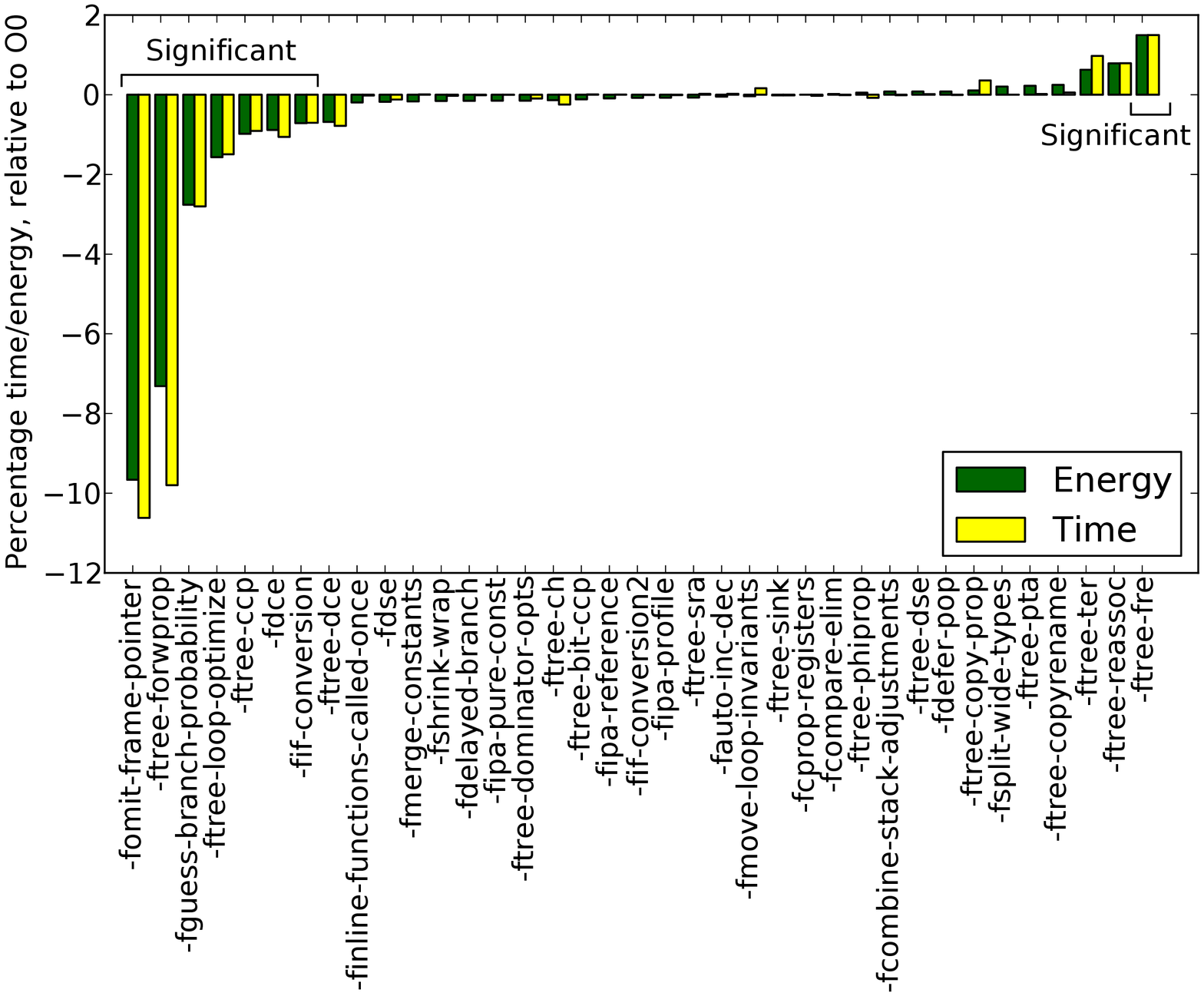}
	\caption{Blowfish benchmark on the Cortex-M0 platform. Individual options enabled at \texttt{O1} are listed.}
	\label{Fig:BlowfishMainEffects}
\end{figure}

\begin{figure}[t!]
	\includegraphics[width=\linewidth,clip,trim=0.5cm 0 2cm 1.8cm]{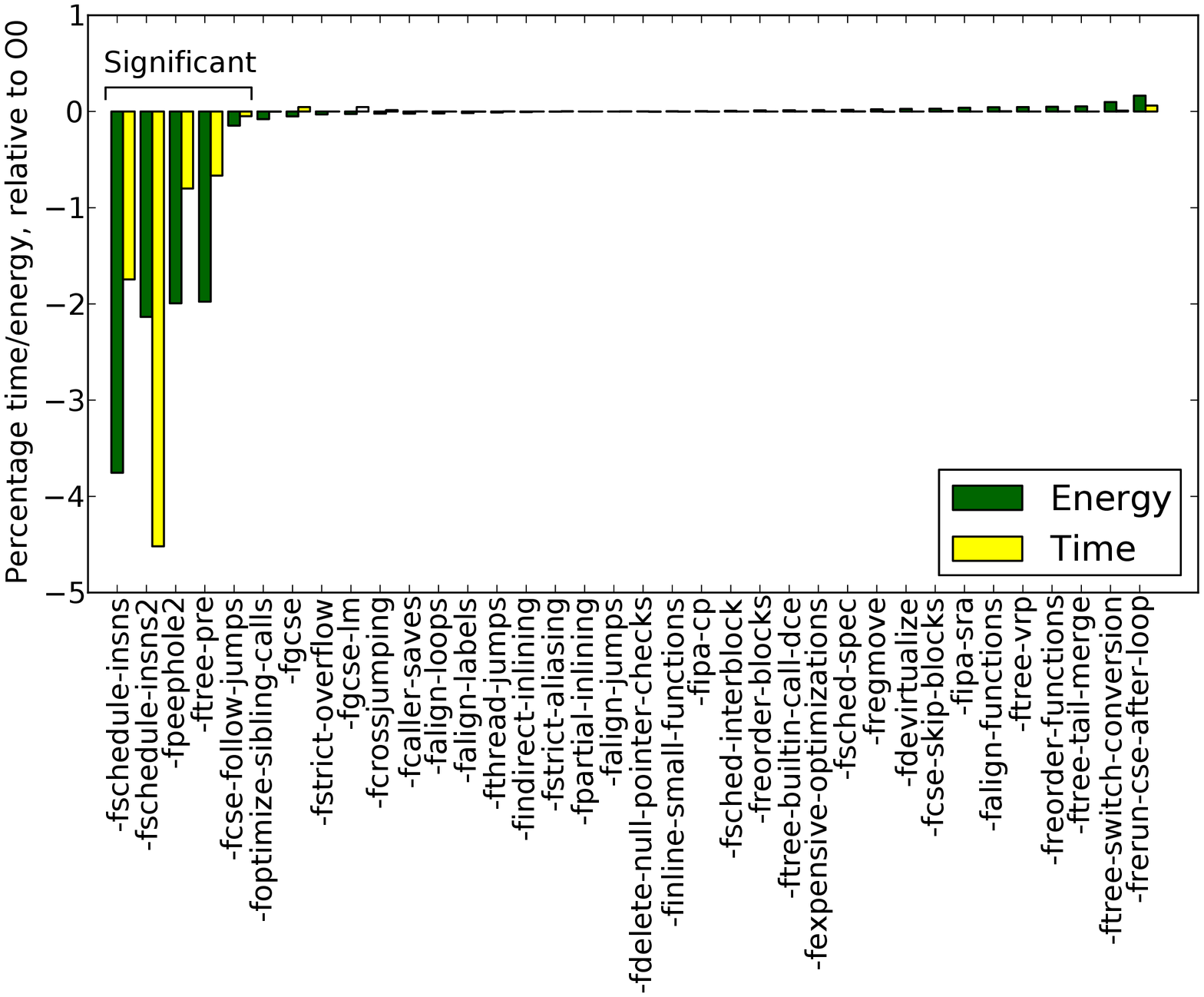}
	\caption{FDCT benchmark on the Cortex-M3 platform. Individual options enabled at \texttt{O2} are listed.}
	\label{Fig:FdctO2MainEffects}
\end{figure}

The results from the FFD experiments provide additional evidence to back up the first hypothesis, that execution time and energy are correlated.

Results showing the correlation between time and energy are shown in Fig.~\ref{Fig:BlowfishMainEffects}. This shows the main effect each optimisation has on the runtime and energy consumption, as calculated by the FFD. A small percentage change is statistically significant because these results are derived from a total of 2048 separate runs. This significance is calculated using the Mann-Whitney test. The bracket above the bars indicates when the result satisfies the following hypothesis: there is 95\% certainty that the result represents a significant impact on the energy consumption of the benchmark.

Fig.~\ref{Fig:FdctO2MainEffects} highlights a discrepancy that occurred between execution time and energy consumption, even for very similar optimisations. The first two options listed (\texttt{-fschedule-insns} and \texttt{-fschedule-insns2}) both schedule instructions to reduce pipeline stalls. However the latter option performs its scheduling pass after register allocation, whereas the first performs it before. The option to schedule instructions after the register allocator can be explained by recognising that the scheduling will reduce stall cycles, which have a below average energy consumption. Overall, this reduces time more than energy (removing cycles that are below average energy will increase the average energy). The other option, however, is more unexpected in that the energy is reduced by a higher proportion than execution time. Upon further investigation this is partly due to fewer spill instructions being generated and partly due to instruction set effects. The scheduling allows causes some register-specific instructions to be converted to ones that are able to access additional registers, further removing the need to access memory.

\subsection{Efficient SIMD Units}

\begin{figure}[t!]
	\includegraphics[width=\linewidth,clip,trim=0.5cm 0 2cm 1.8cm]{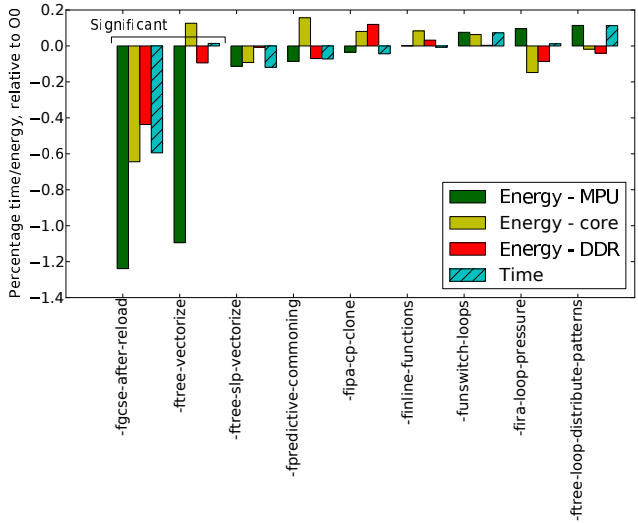}
	\caption{2D FIR benchmark on the Cortex-A8 platform. Individual options enabled at \texttt{O3} are listed.}
	\label{Fig:O3_2dfir_A8}
\end{figure}

\begin{table}
	\centering
	\begin{tabular}{c p{0.3\linewidth} p{0.4\linewidth}}
		\bfseries NEON & \bfseries Instruction Dependencies & \bfseries Continuous Power Consumption \\
		\midrule
		No & \centering Yes & {\hspace{0.85cm} 168 mW } \\
		No & \centering No & {\hspace{0.85cm} 195 mW } \\
		Yes & \centering Yes & {\hspace{0.85cm} 158 mW } \\
		Yes & \centering No & {\hspace{0.85cm} 159 mW } \\
	\end{tabular}
	\caption{Micro-benchmark results for multiplications on the NEON unit, with and without inter-instruction dependencies.}
	\label{Table:SIMD}
\end{table}

In this section we analyse a specific case where energy consumption and execution time are not correlated.

An interesting effect is seen in 2D FIR for the Cortex-A8. The execution time decreases more than the energy consumption up to \texttt{O2}. However, when enabling \texttt{O3} the proportional decrease in energy is greater than execution time (a lower average power). On further investigation, this is caused by the \texttt{-ftree-vectorize} optimisation having an impact on energy consumption with no change in execution time (shown in Fig.~\ref{Fig:O3_2dfir_A8}). This option vectorizes loops, so that SIMD instructions can be inserted. We do not see a performance boost due to the structure of the Cortex-A8 pipeline, where it is expensive to copy results between the NEON unit and the standard registers.

Further investigation of the NEON SIMD unit was done using some simple tests consisting of executing a single instruction many times. The results of these are shown in Tab.~\ref{Table:SIMD}, showing doing continuous multiplication on the NEON unit uses around 20\% less power than using the normal Cortex-A8 multiplier. When considering the similar number of cycles to execute each type of multiply, this results in a reduction in energy consumption when using the NEON unit. This is in line with what previous studies have found~\cite{Ibrahim2009} and shows that by using the hardware to its full capacity, the greatest energy savings can be achieved.


\begin{table*}
	\centering
	\begin{tabular}{l | >{\ttfamily} r >{\ttfamily} c >{\ttfamily} l | >{\ttfamily} r >{\ttfamily} c >{\ttfamily} l | >{\ttfamily} r >{\ttfamily} c >{\ttfamily} l | >{\ttfamily} r >{\ttfamily} c >{\ttfamily} l}
		\bfseries Benchmark &\multicolumn{3}{c}{\rmfamily\bfseries Cortex-M0}
							&\multicolumn{3}{c}{\rmfamily\bfseries Cortex-M3}
							&\multicolumn{3}{c}{\rmfamily\bfseries Cortex-A8}
							&\multicolumn{3}{c}{\rmfamily\bfseries Epiphany} \\
					& $1^{\mathrm{st}}$ & $2^{\mathrm{nd}}$ & $3^{\mathrm{rd}}$ & $1^{\mathrm{st}}$ & $2^{\mathrm{nd}}$ & $3^{\mathrm{rd}}$ & $1^{\mathrm{st}}$ & $2^{\mathrm{nd}}$ & $3^{\mathrm{rd}}$ & $1^{\mathrm{st}}$ & $2^{\mathrm{nd}}$ & $3^{\mathrm{rd}}$ \\
		\midrule
		2dfir          &E&$\cdot$&$\cdot$&  T & G & I      &  N & G & B      &  I &  A & D  \\
		blowfish       &  C & J & E      &  J & C & G      &  K & C & E      &  D &  P & I  \\
		crc32          &F&$\cdot$&$\cdot$&F&$\cdot$&$\cdot$&  F & G &$\cdot$ &  $\cdot$ & $\cdot$ & $\cdot$ \\
		cubic          &  A & H & $\cdot$&  A & H & $\cdot$&A&$\cdot$&$\cdot$&  A &  H & R  \\
		dijkstra       &  H & A & C      &  F & H & A      &  F & H & A      &  H &  A & $\cdot$\\
		fdct           &  J & G & D      &  J & G & K      &  M & K & J      &  A &  I & D  \\
		float\_matmult &  B & E & $\cdot$&  B & E & G      &  N & L &$\cdot$ &  D &  I & A  \\
		int\_matmult   &  B & E & C      &  B & L & F      &  L & N & M      &  A &  I & D  \\
		rijndael       &  C & B & O      &  C & B & O      &  K & C & S      &  D &  K & B  \\
		sha            &  C & B & E      &  B & C & F      &  B & C & M      &  D &  B & Q  \\

	\end{tabular} \\[1em]

	\begin{tabular}{>{\fontsize{9}{1}\selectfont}c | r >{\ttfamily\fontsize{9}{1}\selectfont}l
					>{\fontsize{9}{1}\selectfont}c | r >{\ttfamily\fontsize{9}{1}\selectfont}l }
	\bfseries ID & \multicolumn{2}{l}{\bfseries Count\hspace{1.2cm}Flag} &\bfseries ID & \multicolumn{2}{l}{\bfseries  Count\hspace{1.2cm}Flag} \\
	\midrule
	\fontsize{9}{1}\selectfont
	A & 12 & -ftree-dominator-opts & 	 K & 5  & -fschedule-insns \\
	B & 12 & -ftree-loop-optimise &      L & 3  & -finline-small-functions \\
	C & 11 & -fomit-frame-pointer &      M & 3  & -fschedule-insns2 \\
	D & 8  & -fdce &                     N & 3  & -ftree-pre \\
	E & 7  & -fguess-branch-probability& O & 2  & -ftree-sra \\
	F & 7  & -fmove-loop-invariants &    P & 1  & -fipa-profile \\
	G & 7  & -ftree-ter &                Q & 1  & -ftree-pta \\
	H & 7  & -ftree-fre &                R & 1  & -fcombine-stack-adjustments \\
	I & 6  & -ftree-ch &                 S & 1  & -fgcse  \\
	J & 5  & -ftree-forwprop &           T & 1  & -fpeephole2 \\
	\end{tabular}

	\caption{Table showing the most effective option for each platform-benchmark combination. Options considered were
	optimisations enabled by \texttt{O1}, \texttt{O2} and \texttt{O3} levels.}
	\label{Table:BestFlags}
\end{table*}

\section{The Universality of Flags}
\label{sect:universality_flags}

We have seen large variations based on optimisation flags, and so an interesting problem for compiler designers is how to choose an optimal set of flags across different hardware platforms and applications. This section explores which individual flags had the largest effect in our experiments, our fourth hypothesis: that a consistently good optimisation is not seen across all benchmarks and platforms. Tab.~\ref{Table:BestFlags} lists the results for this section, with the top three optimisation flags (where that optimisation has a significant effect, as per the Mann-Whitney test) identified for each benchmark and platform combination.  Each letter represents an optimisation that is labelled in the table below. We also show the number of times this flag occurs.

Only 20 out of 82 (the number of flags enabled by \texttt{O1}, \texttt{O2} and \texttt{O3}) options examined appear in the table. This supports the argument that many of the options have little effect on the energy consumption, and consequently performance.

For the ARM platforms, a similar set of options appears for the same benchmarks. Common options for the same benchmarks are expected, since optimisations are triggered by the structure of the source code. However, the opposite of this is seen for the Epiphany processor --- there are three optimisations that are consistently effective at reducing energy. A particularly unusual option to be consistently effective is \texttt{-fdce}: dead code elimination, removing code which is never used by the application. However, this also allows the compiler to eliminate parts of the control flow graph, removing branches and decreasing the amount of work the application performs.

The optimisation listed most frequently in the table is \texttt{-ftree-dominator-opts}. The prevalence of this flag is likely due to it enabling several simple optimisation passes, performing optimisations such as copy propagation and expression simplification. Another effective optimisation is \texttt{-fomit-frame-pointer}. This optimisation frees an additional register for general use by not using a frame pointer. This optimisation is seen frequently on the ARM platforms, however not at all on the Epiphany. This is likely due to the ARM processors suffering from greater register pressure since they only have 16 registers compared to the Epiphany's 64.

We see some interesting correlations between platforms. The CRC32 benchmark does not have much optimisation potential since it consists of simple operations in a tight loop. We indeed observe that very few optimisations have a significant effect. Only one common option (\texttt{-fmove-loop-invariants}) appears across three of the four platforms. This optimisation moves redundant calculations out of loops and appears because the CRC32 benchmark has some very tight loops with redundant calculation that can be moved outside the loop regardless of platform.

As observed, some options are seen to affect the energy consumption across benchmarks. This is due to the optimisations being targeted to specific code patterns, which only appear in some of the benchmarks. In particular we see effective options across different ARM platforms, while these same optimisations are not effective on the Epiphany. Since each of the ARM Platforms is using a slightly different instruction set (Thumb, Thumb+Thumb2 and ARM for the Cortex-M0, Cortex-M3 and Cortex-A8 respectively), we infer that the effectiveness of these options is due to commonalities between these instruction sets --- namely the number of registers.

These results show the difficulty of choosing one optimisation which is good in all cases. In many cases the instruction set and micro-architecture of the processor have a large effect on how much the energy consumption is reduced. This means that a singularly good optimisation cannot be chosen.

\subsection{Optimisation Chaos}

In this section, we expand on the theme of there being no universally good optimisation by investigating the effect of interactions between optimisations. We conclude that there is a chaotic relationship between the platform, benchmark, and the effectiveness of the optimisation.

Examining the correlation between optimisations and their effects is a complex issue. Due to non-linear interactions, one would expect that the prediction of effects is difficult. This is borne out by our experimental results: as seen in previous Figs.~\ref{Fig:BlowfishMainEffects} and~\ref{Fig:FdctO2MainEffects}, less than a third of the options have a significant impact. For the other optimisations, higher order interactions cause unpredictable effects, where enabling or disabling a particular optimisation can completely change the effect of many other subsequent optimisation passes.

In Fig.~\ref{Fig:AddsubO1Blowfish}, several unexpected effects worthy of further investigation can be seen. This graph shows individual optimisations being turned on and off, using the \texttt{O1} optimisation level as a base. The flags on the left of the \texttt{O1} section were found to decrease the energy consumption when disabled, an effect not seen in the FFD results. These flags were chosen for further exploration.

To explore this inconsistency, a small case study was performed, where all combinations of four options were explored. The energy figures for exhaustive exploration can be seen in Tab.~\ref{Table:Exhaustive}, with the aim being to ascertain whether the effect of this energy reduction would compound with multiple flags. The \texttt{O1} column of this table shows the results of the options applied over the \texttt{O1} optimisation level. The \texttt{O2} column shows the same but on top of the \texttt{O2} optimisation level.

From the \texttt{O1} column, this it can be seen that there are many interactions occurring between the options, as simply turning all of these options off does not decrease the energy (in fact, it increases the consumption by 1.81\%). Furthermore, when disabled individually, \texttt{-fguess-branch-probability} and \texttt{-ftree-dominator-opts} decrease the energy by 2.49\% and 1.76\% respectively. However, when both are enabled, the energy consumption (relative to \texttt{O1}) is only 0.93\% less, worse than each flag individually.

\begin{table}
	\centering
	\begin{tabular}{c <{\hspace{-2mm}} c <{\hspace{-2mm}} c <{\hspace{-2mm}} c <{\hspace{-2mm}} c r@{.}l c r@{.}l }
		& & & \multicolumn{3}{c}{\null\hfill\bfseries \texttt{O1}} & \multicolumn{3}{c}{\hfill\bfseries \texttt{O2}} \\
		\bfseries X1 & \bfseries X2 & \bfseries X3 & \bfseries X4 & \bfseries (mJ) 	 & \multicolumn{2}{c}{\bfseries (\%)} &
		\bfseries (mJ) 	 & \multicolumn{2}{c}{\bfseries (\%)} \\
		\midrule
		\tY&\tY&\tY&\tY& 5780 & 0&00 	&  5480 &  0&00\\
		\tN&\tY&\tY&\tY& 5640 & -2&49 	&  5540 &  1&00\\
		\tY&\tN&\tY&\tY& 5680 & -1&76 	&  5480 & -0&05\\
		\tN&\tN&\tY&\tY& 5730 & -0&93 	&  5620 &  2&49\\
		\tY&\tY&\tN&\tY& 5650 & -2&28 	&  5490 &  0&09\\
		\tN&\tY&\tN&\tY& 5720 & -0&97 	&  5580 &  1&75\\
		\tY&\tN&\tN&\tY& 5610 & -2&90 	&  5480 &  -0&03\\
		\tN&\tN&\tN&\tY& 5640 & -2&33 	&  5530 &  0&85\\

		\tY&\tY&\tY&\tN& 5760 & -0&34 	&  5460 &  -0&43\\
		\tN&\tY&\tY&\tN& 5720 & -1&09 	&  5480 &  0&03\\
		\tY&\tN&\tY&\tN& 5860 & 1&45 	&  5490 &  0&15\\
		\tN&\tN&\tY&\tN& 5960 & 3&08 	&  5480 &  0&00\\
		\tY&\tY&\tN&\tN& 5890 & 1&91 	&  5470 &  -0&19\\
		\tN&\tY&\tN&\tN& 5870 & 1&61 	&  5570 &  1&57\\
		\tY&\tN&\tN&\tN& 5690 & -1&56 	&  5480 &  -0&03\\
		\tN&\tN&\tN&\tN& 5880 & 1&81 	&  5510 &  0&41\\
	\end{tabular}\\[1em]
	\caption{Exhaustively exploring 4 options compared to \texttt{O1} and \texttt{O2}. (Cortex-M3 with blowfish benchmark). Legend in Tab.~\ref{Table:ExhaustiveLegend}.}
	\label{Table:Exhaustive}
\end{table}
\begin{table}
	\centering
	\begin{tabular}{l p{0.65\linewidth}}
		\bfseries Key & \bfseries Option \\
		\midrule
		X1 & \texttt{-fguess-branch-probability}  \\
		X2 & \texttt{-ftree-dominator-opts}  \\
		X3 & \texttt{-ftree-ch} \\
		X4 & \texttt{-fif-conversion} \\
		Abs (mJ) & Absolute energy measurement in millijoules. \\
		\texttt{O1} (\%) & Percentage relative to \texttt{O1}. \\
		\texttt{O2} (\%) & Percentage relative to \texttt{O2}. \\
		\tY & Optimisation is enabled.  \\
		\tN & Optimisation is disabled.  \\
	\end{tabular}
	\caption{Legend for Tab.~\ref{Table:Exhaustive}.}
	\label{Table:ExhaustiveLegend}
\end{table}

Different results are seen entirely in the \texttt{O2} column, with options that decreased energy consumption on top of \texttt{O1} have little or the opposite effect when applied on top of \texttt{O2}.

This unpredictability suggests that these options have many interdependencies that are difficult to predict up front. It also makes choosing an optimal set of optimisations very challenging. Therefore, one of our findings is that it is very unlikely that any accurate prediction mechanism for considering an optimisation and its effect on a target system exists: the effect will always be highly dependent on the application to be used and the platform upon which it resides.

\section{What does this mean for the application developer?}

The existing collections of optimisations at the various levels do a good job of optimising for performance, and consequently, energy. These strike a good balance between ease of use and performance. However, they will never be as effective as those generated by searching through the full optimisation space. To avoid running many tests to find a good solution, developing machine-learning compiler technologies similar to MILEPOST GCC~\cite{Fursin2011} would be fitting. A reasonable set of optimisations can be predicted based on high-level features and an architecture selection, and this would greatly reduce the time spent searching as demonstrated by MILEPOST GCC. This is especially true as the effectiveness and type of optimisation was found to be heavily based on the platform and the structure of the application being compiled (Sect.~\ref{sect:universality_flags}). Predicting the optimisations in this way would reduce compile times as well as the energy and execution time of the application.

This study focused on GCC, since it is a mature compiler supporting many different platforms and optimisations. As an alternative, the LLVM compiler~\cite{LLVM} is relatively new, with a well defined set of optimisation passes, whose order can easily be specified. This extra flexibility means there may be a better solution to find, but also that it is essentially searching for a sharper needle in a bigger haystack~\cite{Kulkarni2007}. The benefits from having this much larger space to explore may not be worth the trade-off of the time it takes to find it.

\section{What does this mean for the compiler writer?}

When designing a new optimisation, a compiler writer must check whether the optimisation is effective, and under what conditions. Using fractional factorial, design a compiler writer can check whether the pass is effective when combined with an arbitrary set of other optimisations. This avoids the case of the optimisation being tested in isolation, which will result in an incorrect analysis because of the interactions between optimisations. We would recommend that, when selecting optimisations to be included in a broad optimisation level, the optimisation is evaluated in this way and only selected if it has a non-negative effect over all of the benchmarks.

All the optimisations we show in this paper are designed for either performance or code size. This means we cannot draw conclusions about the effect of dedicated compiler optimisations targeting energy such as those shown in the related work (Sect.~\ref{subsect:energyoptimisations}). Although all optimisation targets may be beneficial for energy usage, dedicated energy flags would have to compete against these other optimisation metrics, meaning that even if they operate well in isolation, they may not do well when grouped. There are many opportunities for further work in this area.

\section{Conclusion}

The first hypothesis of energy consumption and execution time being correlated in the general case was found to be correct across many platforms and benchmarks. This was first shown to be true by the high level results, showing only the overall optimisation levels. The more detailed fractional factorial design runs also demonstrated this result, showing that most optimisations had the same relative effect on energy and time. This result occurs because the majority of optimisations focus on reducing the total amount of work performed by the benchmarks --- thus minimising both energy consumption and execution time.

By adding and subtracting individual flags on top of the whole optimisation levels we have shown that a better set of flags exists, which can produce more optimal applications. This validates our second hypothesis, giving results in line with much previous work.

The third hypothesis stated that it was possible to efficiently search the optimisation space to gain information about the effectiveness of each optimisation. To perform this we leveraged fractional factorial designs, allowing us to test each optimisation in a greatly reduced number of runs. This method allowed us to explore complex effects seen on the Cortex-A8, where the SIMD unit helped achieve lower energy consumption.

The fourth hypothesis of there being no optimisation which was effective for all benchmarks and platforms was evaluated using fractional factorial designs. We were able to extract the most effective optimisations for each benchmark and platform pair and these results showed that there was no single optimisation that was universally effective. Further analysis of adding and subtracting individual flags showed that the optimisation space is chaotic, with optimisations interacting in unpredictable ways.

The compiler writer can use these results and the fractional factorial design method to evaluate potential optimisation passes, ensuring that they perform well in a variety of configurations. Until a method for resolving the interactions between optimisations is found, it is envisioned that the developer could use this technique to eliminate optimisations that are not having a positive effect on their application. This will speed up compilation time as well as potentially improving the performance of their application.

\ack

This study was funded by Embecosm. The original research proposal was a result of the Energy Aware COmputing (EACO) workshops at the University of Bristol, sponsored by the Institute for Advanced Studies. The first author was  partly sponsored by EPSRC's Doctoral Training Account EP/K502996/1.

\appendix

\section{Hardware Setup}

All the measurements were taken using the INA219 power monitoring IC~\cite{INA219}, which provides power, current and voltage outputs.

The Cortex-M0 and Cortex-M3 boards both have a single measurement point, recording the power consumed by the whole microprocessor. For the BeagleBone there are three available measurement points: the Cortex-A8 core (including caches), on-chip peripherals (power management, bus controllers) and the external SDRAM memory IC. This allows the effect of the compiler optimisations on the memory to be recorded. Adapteva's Epiphany board has two measurement points: the core power consumption and IO power consumption, whereas the XMOS board's measurement point gathers power consumption data for the core of the processor.

The hardware measurements have several sources of error. The most apparent errors are variations in the timing: the INA219 is sampled at intervals of 1~ms and the power measurement integrated over this. Small inaccuracies occur from jitter in this interval. The ADC in the INA219 also fluctuated by $\pm$30~$\mathrm{\mu V}$, however this was close to the noise floor of the measurements, so had no significant effect on the results.


\begin{thebibliography}{99}

\bibitem{SmartPhonePower}
Carroll, A. and Heiser, G. (2010) An analysis of power consumption in a smartphone.
{\em Proc. USENIX}, Boston, MA, USA, 22--25 June, pp. 21--21. USENIX Association Berkeley, CA, USA.

\bibitem{Lofikamran2012}
Lotfi-kamran, P., Grot, B., Ferdman, M., Volos, S., and Kocberber, O. (2012) {Scale-Out Processors}.
{\em Int. Symp. Computer Architecture 12}, Portland, Oregon, 9--13 June, pp. 500--511. IEEE Computer Society, Washington, DC, USA.

\bibitem{Hollis2012}
Hollis, S. J., Jackson, C., Bogdan, P., and Marculescu, R. (2012) {Exploiting Emergence in On-chip Interconnects}.
{\em IEEE Transactions on Computers}, {\bf PP(99)},
  1--14.

\bibitem{Pan2006a}
Pan, Z. and Eigenmann, R. (2006) {Fast and effective orchestration of compiler optimizations for automatic performance tuning}.
{\em Int. Symp. Code Generation and Optimization 06}, New York, USA, 26--29 March, pp. 319--332. IEEE Computer Society, Washington, DC, USA.

\bibitem{Haneda2005}
Haneda, M., Knijnenburg, P. M.~W., and Wijshoff, H. A.~G. (2005) {Automatic selection of compiler options using non-parametric inferential statistics}.
{\em Proc. Int. Conf. Parallel Architectures and Compilation Techniques}, St. Louis, USA, 17--21 September, pp. 123--132. IEEE Computer Society, Washington, DC, USA.

\bibitem{Lin2008}
Lin, S. C., Chang, C. K., and Lin, N. W. (2008) {Automatic selection of GCC optimization options using a gene weighted genetic algorithm}.
{\em Proc. Computer Systems Architecture Conference}, Hsinchu, Taiwan, 4--6 August, pp. 1--8. IEEE Computer Society, Washington, DC, USA.

\bibitem{Kisuki1999}
Kisuki, T., Knijnenburg, P. M.~W., O'Boyle, M. F.~P., Bodin, F. and Wijshoff, H. A. G. (1999) {A feasibility study in iterative compilation}.
{\em Int. Symp. High Performance Computing}, Kyoto, Japan, 26--28 May, pp. 121--132. Springer Berlin Heidelberg.

\bibitem{Purini2013}
Purini, S. and Jain, L. (2013) {Automatic selection of compiler options using non-parametric inferential statistics}.
{\em Transactions on Architecture and Code Optimization}, {\bf 9}, 1--23. ACM, New York, USA.

\bibitem{EffectOfCompilerOptimisationsOnPentium4}
Seng, J.~S. and Tullsen, D.~M. (2003) The effect of compiler optimizations on pentium 4 power consumption.
{\em Proc. Workshop on Interaction between Compilers and Computer Architectures},  Anaheim, CA, USA, 8 Feb, pp. 51. IEEE Computer Society, Washington, DC, USA.

\bibitem{Ibrahim2009b}
Ibrahim, M.~E.~A., Rupp, M. and Habib, S. E.-D. (2009) Compiler-based optimizations impact on embedded software power consumption
{\em Workshop on Circuits and Systems and TAISA Conference}, Toulouse, France, 28 June -- 1 July, pp. 1--4. IEEE.

\bibitem{CompilingForPerformancePower}
Valluri, M. and John, L. (2001).
{\em Is compiling for performance == compiling for power?}

\bibitem{Wattch}
Brooks, D., Tiwari, V., and Martonosi, M. (2000) Wattch: a framework for architectural-level power analysis and optimizations.
{\em Int. Symp. Computer Architecture}, Vancouver, BC, Canada, 14 June, pp. 83--94. IEEE Computer Society, Washington, DC, USA.

\bibitem{Ayala2004}
Ayala, J. and L\'{o}pez-Vallejo, M. (2004) {Improving register file banking with a power-aware unroller}.
{\em Proc. PARC} Pisa, Italy. 15--20.

\bibitem{Zhu2004}
Zhu, Y., Magklis, G., and Scott, M. (2004) {The energy impact of aggressive loop fusion}.
{\em Proc. Int. Conf. Parallel Architectures and Compilation Techniques}. Antibes Juan-les-Pins, France, 29 Sept -- 3 Oct, pp. 153--164. IEEE Computer Society, Washington, DC, USA.

\bibitem{Kim2012}
Kim, B., Cho, Y., and Hong, J. (2012) {An Efficient Function Inlining Scheme for Resource-Constrained Embedded Systems}.
{\em Journal of Information Science and Engineering}, {\bf 28}, 859--874.

\bibitem{Toburen1998}
Toburen, M., Conte, T., and Reilly, M. (1998) {Instruction scheduling for low power dissipation in high performance microprocessors}.
{\em Proc. Power Driven Microarchitecture Workshop}, Barcelona, Spain, 28 June, pp. 14--19.

\bibitem{SimplePower}
Ye, W., Vijaykrishnan, N., Kandemir, M., and Irwin, M.~J. (2000) The design and use of simplepower: a cycle-accurate energy estimation tool.
{\em Proc. Design Automation Conference}, Los Angeles, California, USA, 5--9 June, pp. 340--345. ACM.

\bibitem{Steinke2001}
Steinke, S., Knauer, M., Wehmeyer, L., and Marwedel, P. (2001) {An accurate and fine grain instruction-level energy model supporting software optimizations}.
{\em Proc. PATMOS}, Yverdon-Les-Bains, Switzerland, 26--28 Sept.

\bibitem{Qu2000}
Qu, G., Kawabe, N., Usami, K., and Potkonjak, M. (2000) {Function-level power estimation methodology for microprocessors}.
{\em Proc. Design Automation Conference}, Los Angeles, CA, USA, 5--9 June, pp. 810--813. ACM.

\bibitem{Blume2007}
Blume, H., Becker, D., Rotenberg, L., Botteck, M., Brakensiek, J., and Noll, T. (2007) {Hybrid functional- and instruction-level power modeling for embedded and heterogeneous processor architectures}.
{\em Journal of Systems Architecture}, {\bf 53}, 689--702.

\bibitem{CompilationTechniquesForLowEnergy}
Tiwari, V., Malik, S., and Wolfe, A. (1994) Compilation techniques for low energy: an overview.
{\em IEEE Symp. Low Power Electronics}, San Diego, CA, USA, 10--12 Oct, pp. 38 --39. IEEE Computer Society, Washington, DC, USA.

\bibitem{Seth2001}
Seth, A., Keskar, R.~B., and Venugopal, R. (2001) {Algorithms for energy optimization using processor instructions}.
{\em Proc. Int. Conf. Compilers, Architecture, and Synthesis for Embedded Systems}, Atlanta, Georgia, USA, pp. 195--202. ACM.

\bibitem{Ibrahim2009}
Ibrahim, M. E.~A., Rupp, M., and Fahmy, H. A.~H. (2009) {Code transformations and SIMD impact on embedded software energy/power consumption}.
{\em Proc. Int. Conf. Computer Engineering \& Systems},  Cairo, Egypt, 14--16 Dec, pp. 27--32. IEEE Computer Society, Washington, DC, USA.

\bibitem{Parikh}
Parikh, A., Kandemir, M., Vijaykrishnan, N., and Irwin, M. (2000) {Instruction scheduling based on energy and performance constraints}.
{\em Proc. IEEE Computer Society Workshop on VLSI}, Orlando, FL, USA, pp. 37--42. IEEE Computer Society, Washington, DC, USA.

\bibitem{WhatCanAPoorCompilerDo}
Chakrapani, L.~N. and et al. (2001) The emerging power crisis in embedded processors: What can a (poor) compiler do?
{\em Proc. Int. Conf. Compilers, Architecture, and Synthesis for
  Embedded Systems}, Atlanta, Georgia, USA, pp. 176--180. ACM.

\bibitem{Steinke}
Steinke, S., Wehmeyer, L. and Marwedel, P. (2002) {Assigning program and data objects to scratchpad for energy reduction}.
{\em Proc. Design Automation and Test in Europe}, Paris, France, 4--8 March, pp. 409--415. IEEE.

\bibitem{Cao2001}
Cao, Y. and Yasuura, H. (2001) {A system-level energy minimization approach using datapath width optimization}.
{\em Proc. Int. Symp. Low Power Electronics and Design}, California, USA, 6--7 August, pp. 231--237. IEEE.

\bibitem{IterativeCompilationForEnergy}
Gheorghita, S.~V., Corporaal, H., and Basten, T. (2005) Iterative compilation for energy reduction.
{\em J. Embedded Computing}, {\bf 1}, 509--520.

\bibitem{Almagor2004}
Almagor, L., Cooper, K.~D., and Grosul, A. (2004) {Finding effective compilation sequences}.
{\em Proc. ACM Conf. Languages, Compilers, and Tools for Embedded Systems}, Washington, DC, USA, 11--13 June, pp. 231--239. ACM.

\bibitem{Lokuciejewski2011}
Lokuciejewski, P., Plazar, S., Falk, H., Marwedel, P. and Thiele, L. (2011) {Approximating Pareto optimal compiler optimization sequences --- a trade-off between WCET, ACET and code size}.
{\em Software --- Practice and Experience}, {\bf 41}, 1437--1458.

\bibitem{BoxHunter}
George E. P.~Box, J. S.~H., William G.~Hunter (1978) {\em {Statistics for Experimenters: An Introduction to Design, Data Analysis, and Model Building}}. John Wiley \& Sons, New York.

\bibitem{IntelPaper}
Chow, K. and Wu, Y. (1999) Feedback-directed selection and characterization of compiler optimizations.
{\em Workshop on Feedback Directed Optimization}, Austin, Texas, USA, 1 Dec.

\bibitem{Haneda2005b}
Haneda, M. Knijnenburg, P. M. W. and Wijshoff, H. A. G. (2005) {Optimizing General Purpose Compiler Optimization}.
{\em Proc. 2nd Conf. Computing Frontiers}, Ischia, Italy, 4--6 May, pp. 180--188. ACM.

\bibitem{EnergyReductionCompilerOptions}
Patyk, T., Hannula, H., Kellomaki, P., and Takala, J. (2009) {Energy consumption reduction by automatic selection of compiler options}.
{\em Proc. Int. Symp. Signals, Circuits and Systems}, Iasi, Romania, 9--10 July, pp. 1--4. IEEE Computer Society, Washington, DC, USA.

\bibitem{Fursin2011}
Fursin, G., Kashnikov, Y., and Memon, A.~W. (2011) {Milepost GCC: machine learning enabled self-tuning compiler}.
{\em Int. J. Parallel Programming}, {\bf 39}, 296--327.

\bibitem{BenchmarkPaper}
Pallister, J., Hollis, S. and Bennett, J. (2013) {BEEBS: Open benchmarks for energy measurements on embedded platforms}.
[Preprint] Available from: http://www.cs.bris.ac.uk/Research/Micro/beebs.jsp [Accessed 23rd August 2013].

\bibitem{MiBench}
Guthaus, M.~R., Ringenberg, J.~S., Ernst, D., Austin, T.~M., Mudge, T., and Brown, R.~B. (2001) Mibench: A free, commercially representative embedded benchmark suite.
\newblock {\em Proc. IEEE Workshop on Workload Characterization},  Washington, DC, USA, pp. 3--14. IEEE Computer Society, Washington, DC, USA.

\bibitem{WCET}
Gustafsson, J., Betts, A., Ermedahl, A., and Lisper, B. (2010) The m{\"a}lardalen wcet benchmarks - past, present and future.
\newblock {\em Proc. 10th Int. Workshop on Worst-Case Execution Time Analysis}, Brussels, Belgium, 6 July, pp. 137--147.

\bibitem{Hennessy2012}
Hennessy, J.~L. and Patterson, D.~A. (2012) {\em {Computer Architecture: A Quantitative Approach}}, 5th edition. Morgan Kaufmann, MA, USA.

\bibitem{GCC} {Free Software Foundation}, GCC, the GNU Compiler Collection, {http://gnu.gcc.org/}, (Accessed 2013/03/20).

\bibitem{Yiu2010}
Yiu, J. (2010) {\em {The Definitive Guide to the ARM Cortex-M3}},  2nd edition. Newnes, MA, USA.

\bibitem{LLVM}
{University of Illinois}. {The LLVM Compiler Infrastructure}, {http://llvm.org/}, (Accessed 2013/03/20).

\bibitem{R}
{Free Software Foundation}, {The R Project for Statistical Computing}, {http://www.r-project.org/}, (Accessed 2013/03/20).

\bibitem{FrF2}
Groemping, U. and Groemping, M.~U. (2012).
{FrF2: Fractional Factorial designs with 2-level factors}.

\bibitem{Kulkarni2007}
Kulkarni, Prasad A., Whalley, David B., Tyson, Gary S. and Davidson, Jack W. (2007) {Evaluating heuristic optimization phase order search algorithms}.
{\em Int. Symp. Code Generation and Optimization}, California, USA, 11--14 March, pp. 157--169. IEEE.

\bibitem{INA219}
Texas Instruments. (2011) {\em INA219: Current / Power Monitor with I2C$\textsuperscript{TM}$ Interface}. Texas Instruments, Dallas, Texas, USA.

\end{thebibliography}
\end{document}